\documentclass[14pt]{article}
\pdfoutput=1
\usepackage{amsmath,amssymb,amsfonts,amsthm,bm,bbm,cancel,wasysym}
\usepackage{epsfig,graphics,graphicx,epstopdf}
\usepackage{array,booktabs,colortbl,colordvi,multirow}
\usepackage{colordvi,color,xcolor}
\usepackage{hyperref}
\usepackage{rotating}

\usepackage{verbatim}
\usepackage{cite}
\usepackage{subfig}
\usepackage{setspace}
\usepackage{url}
\usepackage[percent]{overpic}
\usepackage{slashed}
\usepackage{authblk}
\usepackage{xspace}
\usepackage{fullpage}
\usepackage{hyperref}

\def\ie{{\it i.e.}}
\def\eg{{\it e.g.}}
\def\etc{{\it etc}}

\newcommand{\kslash}{\cancel{k}}
\newcommand{\pslash}{\cancel{p}}

\def\to{\rightarrow}

\newskip\zatskip \zatskip=0pt plus0pt minus0pt
\def\matth{\mathsurround=0pt}
\def\lsim{\mathrel{\mathpalette\atversim<}}
\def\gsim{\mathrel{\mathpalette\atversim>}}
\def\atversim#1#2{\lower0.7ex\vbox{\baselineskip\zatskip\lineskip\zatskip
  \lineskiplimit 0pt\ialign{$\matth#1\hfil##\hfil$\crcr#2\crcr\sim\crcr}}}

\begin{document}


\begin{flushright}
SLAC-PUB-260817\\
\today
\end{flushright}
\vspace*{5mm}

\renewcommand{\thefootnote}{\fnsymbol{footnote}}
\setcounter{footnote}{1}

\begin{center}

{\Large {\bf Invisible Decays of the $Z$ in Portal Matter Models at Tera-Z}}\\

\vspace*{0.75cm}

{\bf Thomas G. Rizzo}~\footnote{rizzo@slac.stanford.edu}

\vspace{0.5cm}

{SLAC National Accelerator Laboratory}\\ 
{2575 Sand Hill Rd., Menlo Park, CA, 94025 USA}

\end{center}
\vspace{.5cm}


\begin{abstract}
\noindent  
Searches for new contributions to the invisible width of the $Z$, $\Delta\Gamma_{inv}$, can be used to provide model-dependent constraints on new physics that may arise from an 
unknown dark sector. In the vector portal setup considered here, after a survey of several tree-level scenarios, it is shown that loops of portal matter (PM) fields, having both Standard Model (SM) 
and dark sector quantum numbers, can naturally lead to new contributions to $\Delta\Gamma_{inv}$ that satisfy current experimental constraints while also being simultaneously accessible 
to measurements at the future Tera-Z run of the FCC-ee. Such PM fields, which here take the form of heavy vector-like fermions in SM-like representations, are already well known to be able to 
induce kinetic mixing (KM) between the SM and dark photons (DP) at low energies. This analysis is 
performed within a simplified version of a previously examined `$E_6-$like' setup that also predicts a calculable value for the strength of this KM in a phenomenologically interesting range. On the 
other hand, owing to the constraint on the invisible branching fraction of the Higgs from the LHC, the tree-level mixing of the SM and a weak isosinglet dark Higgs fields yields a value of 
$\Delta\Gamma_{inv}$ which is far too small to be observable even by Tera-Z. A similar result is found to hold for the shift in the $Z$'s invisible width induced by the mass and/or KM of the $Z$ 
with the light DP which then subsequently decays to dark matter (DM). On the other hand, PM models wherein the light dark Higgs carries a significant admixture of a $T_3\neq 0$ weak isospin 
component are shown to already violate existing $\Delta\Gamma_{inv}$ experimental constraints in the absence of parameter fine tuning. Future measurements of $\Delta \Gamma_{inv}$ can 
thus play an important role in providing indirect constraints on dark sector interactions and PM model building, in particular, on PM loops not necessarily directly associated with the generation of 
the effective KM of the DP with the corresponding SM field. 
\end{abstract}

\vspace{0.5cm}
\renewcommand{\thefootnote}{\arabic{footnote}}
\setcounter{footnote}{0}
\thispagestyle{empty}
\vfill
\newpage
\setcounter{page}{1}



\section{Introduction and PM Model Background}

The nature of dark matter (DM) is one of the major mysteries left standing after the many successes of the Standard Model (SM), with many associated questions left to answer: what is it, how did 
it come to exist, why is there as much of it\cite{Planck:2018vyg} and does it interact with us, \ie, the SM  in any way other than through gravity? There is no shortage of models in the 
literature that attempt to 
address at least some of these basic questions about DM. Of these, an attractive possibility which we will focus on below is that DM is a light ($\lsim 1$ GeV) thermal relic which talks to the SM via 
the kinetic mixing (KM)/vector boson portal\cite{KM,vectorportal,Gherghetta:2019coi}.  In its simplest form, this setup posits the existence of a new dark $U(1)_D$ gauge group having a gauge 
coupling $g_D$,  corresponding to a dark charge, $Q_D$, and a new massive gauge field (here also $\lsim 1$ GeV) called the dark photon (DP),  
$V$\cite{Fabbrichesi:2020wbt,Graham:2021ggy,Barducci:2021egn,Alonso-Gonzalez:2025xqg,Jorge:2026bbs,Caputo:2026pdw}. By definition, the DM carries a non-zero value of $Q_D$ but all of the 
SM fields (fermions, Higgs and gauge bosons) by contrast will have $Q_D=0$ and so their tree-level couplings to the DP are absent. Here, it will be assumed that the mass of the DP is generated in an 
a manner similar to the spontaneous symmetry breaking in the SM, \ie, via the vev(s) of one or more dark Higgs fields, $h_D$. In the simplest setup, the dark Higgs is usually taken to be just a 
SM singlet but in more complex scenarios, as we will consider below, one or more dark Higgs fields may also transform non-trivially under the SM gauge symmetries\cite{Li:2024wqj}. Of course in 
such more general setups it is possible that there exists an entire class of particles (fermions and/or bosons) which will carry {\it both} the SM as well as $U(1)_D$ quantum numbers which we 
have referred to as portal matter (PM) in our earlier work\cite{Rizzo:2018vlb,Rueter:2019wdf,Kim:2019oyh,Rueter:2020qhf,Wojcik:2020wgm,Rizzo:2021lob,Rizzo:2022qan,Wojcik:2022rtk,Rizzo:2022jti,Rizzo:2022lpm,Bauer:2022nwt,Wojcik:2022woa,Carvunis:2022yur,Verma:2022nyd,Rizzo:2023qbj,Wojcik:2023ggt,Rizzo:2023kvy,Rizzo:2023djp,Rizzo:2024bhn,Ardu:2024bxg,Rizzo:2024kzu,Rizzo:2025tap,
Tewary:2025vij,Rizzo:2025fsy,Rizzo:2026dsj,Rizzo:2026mza}. 

It is easily imagined that such PM fields will quite naturally also be present in generalizations of the above simple KM setup wherein $U(1)_D$ is embedded into some larger, \ie, non-abelian gauge 
structure, $G_D$, which we imagine breaking at the $\sim 10$ TeV scale and in doing so also generating the masses for the PM.  As might be expected, if these PM fields are fermionic, then they 
are required to be vector-like (VL) with respect to the SM gauge group, $G_{SM}=SU(3)_c\times SU(2)_L \times U(1)_Y=3_c2_L1_Y$, in order to avoid a multitude of well-known unitarity, 
precision electroweak and Higgs width/production cross section constraints\cite{CarcamoHernandez:2023wzf,CMS:2024bni,Alves:2023ufm,Banerjee:2024zvg,Guedes:2021oqx,Adhikary:2024esf,Benbrik:2024fku,Albergaria:2024pji,Chen:2017hak,
Biekotter:2016kgi}. 
More generally, one observes that these scalar and/or fermionic PM fields will play a rather special role in KM setups: depending upon the specifics of their SM transformation 
properties, loops of PM particles will end up sourcing potentially sizable KM between the DP and the neutral SM $SU(2)_L$ gauge field, $W_3$, and/or $B$,  the $U(1)_Y$ gauge field coupling to 
hypercharge, via 1-loop vacuum polarization-like graphs. These loops will then appear at the lower energies far below the weak scale as the familiar KM of the SM photon and $V$.  
As we assume here, when the DP and DM are both in the $\lsim 1$ GeV mass range the strength of this PM loop-induced KM, $\epsilon$, can be {\it calculated} in more UV-complete scenarios 
to be in the range necessary to allow the DM to have its observed relic abundance via the usual thermal processes, as in the case of the familiar 
WIMPS\cite{Arcadi:2017kky,Roszkowski:2017nbc,Arcadi:2024ukq}, while also satisfying other numerous experimental constraints.  This happens because, due to this KM, the $Q_D=0$ charged 
fields of the SM will pick up a small loop-induced coupling to the DP with a strength $e\epsilon Q_{em}$, with $e$ being the usual charge of the proton.  
It is to be noted that in the simplest versions of this setup, no {\it other} additional interactions between the DP (and hence DM) with the SM are expected to exist. However, again depending upon 
the SM quantum numbers of the PM fields, we have shown in earlier work that other new interactions beyond KM can be generated at loop-level, occasionally in the form of dark 
moments{\footnote{ Of course, further distinctive interactions, such as those discussed below, are also possible.}} This 
happens most frequently when the $U(1)_D$ is embedded in some larger dark gauge group, $G_D$, as mentioned above. Since $G_D$ needs to break at the $\sim 10$ TeV scale, say, 
an enlarged dark Higgs sector is also needed in this case to generate the relevant PM masses and to break the symmetries as required.  In such contexts it is natural to also explore the possibility 
of a more unified treatment of the SM and the dark sectors; a somewhat simplified, `toy model' version of one such scenario\cite{Rueter:2019wdf}, inspired by the 
group $E_6$\cite{Hewett:1988xc}, and which we have exploited in several of the earlier works above, will be further employed in the analysis below.

 As noted, depending upon their quantum numbers, PM fields can generate new interactions which can be searched for in present and future experiments. While the direct searches for the heavy 
 PM fields themselves require the highest energy colliders, new mixing/loop-induced interactions involving more easily produced SM states can offer additional windows\cite{Rizzo:2026mza} 
 into the dark sector at somewhat lower energies. One such possibility is the subject of the present paper: here we consider the sensitivity of the partial decay width for $Z\to {\rm invisible}$, 
 $\Gamma_{inv}$, as an indirect probe for such PM interactions and of dark sector 
 physics more generally. This process, of course, already occurs in the SM via the decay of the $Z$ into the 3 generations of neutrinos with a rate experimentally determined to be very close to what is 
 predicted by the SM\cite{limit} and so any additional contributions from new physics sources cannot be too large thus leading to constraints on the new physics parameter space. 
 This SM prediction will be much more definitively tested than presently by improved measurements having 
 essentially 2 orders of magnitude more in sensitivity, down to the level of $\Delta \Gamma_{inv}\simeq 25$ keV, in the not too distant future after the turn on of the FCC-ee running in the Tera-Z 
 mode\cite{limit}.  In the current PM setup, various new sources for this $Z$ decay mode can exist which now involve the light DM, DP and/or dark Higgs final states through a variety of 
 available mechanisms. Here we will explore several of these possibilities: in particular we show that two of these prospects, induced by Higgs or gauge boson mixing, lead to rates far smaller 
 than those accessible to Tera-Z, while in a third case, also involving mixings within an enlarged dark Higgs sector, some parameter fine-tuning is required even to satisfy the present bound of 
 roughly $\Delta \Gamma_{inv}\lsim 2.4$ MeV\cite{limit}. As a final possibility, we consider the PM triangle loop-induced process $Z\to h_DG_V(\simeq V_L)$, where $G_V$ is the Goldstone boson 
 of the DP which is essentially its longitudinal component, $V_L$,  in a simplified version of the `$E_6$-inspired' model mentioned above where the PM fields transform as, in $SU(5)$ language, an 
 anomaly-free ${\bf 5}+\bar {\bf 5}$ pair of vector-like fermions.  We will show that for typical choices of the four model parameters, \ie, two PM fermion masses at the TeV scale plus two O(1) 
 Yukawa couplings, this setup generally predicts values of $\Delta \Gamma_{inv}$ which are both safely below the current bound yet lying within the range that is accessible to the future 
 measurements possible at Tera-Z. 
 
The outline of this paper is as follows:  After this Introduction, in Section 2, we will discuss a set of PM models wherein $\Delta \Gamma_{inv}$ is induced via the mixings among various new 
dark states with the SM and/or with each other. Three cases are considered:  A first example of this is the familiar mass mixing between the SM and an isosinglet dark Higgs field via quartic terms in 
the scalar potential which then allows for the decay $Z\to h_D \bar \nu_i \nu_i$ with $h_D$ decaying into DP pairs. However, due to constraints from SM Higgs invisible decays, we demonstrate that 
the rate for this process is necessarily far below what Tera-Z can probe. A second possibility is also familiar, \ie, the mass and/or KM of the DP with the $Z$ allowing for the $Z$ to decay directly 
to DM. Again, we show that this rate is far too small to be seen due to both the expected size of the ratio $m_V^2/m_Z^2 <10^{-4}$ and the required size of $\epsilon \sim 10^{-4}-10^{-3}$ needed 
to satisfy all the experimental 
constraints.  A third possibility is the mixing of two Higgs fields, both of which carry non-zero values of $Q_D$ and at least one of which also carries a non-zero third component of weak isospin, \ie, 
$T_3\neq 0$, which mix to yield the light dark Higgs state, $h_D$, along with accompanying heavy scalars. This setup is immediately excluded due to its large contribution to $\Delta \Gamma_{inv}$ 
which exceeds the current bound by a substantial factor unless the mixing between the dark Higgs fields to form $h_D$ is tuned so that it is mostly composed of a $T_3=0$ state.
In Section 3, we turn to the loop-induced process $Z\to h_DG_V$ which occurs via the couplings of the PM fields themselves to the $Z$, as they have $Q_{em},T_3 \neq 0$, as well as their 
couplings to the dark sector fields $h_D,V$ as they also carry $Q_D\neq 0$ as well. As noted above, these calculations are performed within a truncated `$E_6$-inspired' class of models employed 
in our earlier work but the analysis is seen to be easily generalized to other more complex cases. We show that this process can yield a rate potentially observable at Tera-Z while also 
satisfying existing constraints and leads to a calculable value for $\epsilon$ in the experimentally interesting range.  Finally, a discussion of our results, possible future directions,  and our 
conclusions are presented in Section 4.


\section{$Z$ Couplings to the Dark Sector via Mixing}

In order for the PM model, or more generally the KM/vector portal model, to generate new contributions to the invisible width of the $Z$, $\Delta \Gamma_{inv}$, that can be accessible to Tera-Z,  
some tree level couplings/mixings between states and/or new loop processes must be present. We first turn to a brief discussion of some of these tree-level possibilities within this overall context.

\subsection{SM - Dark Higgs Mixing}

\begin{figure}[htbp] 
\hspace*{1.5cm}\centerline{\includegraphics[width=5.5in,angle=0]{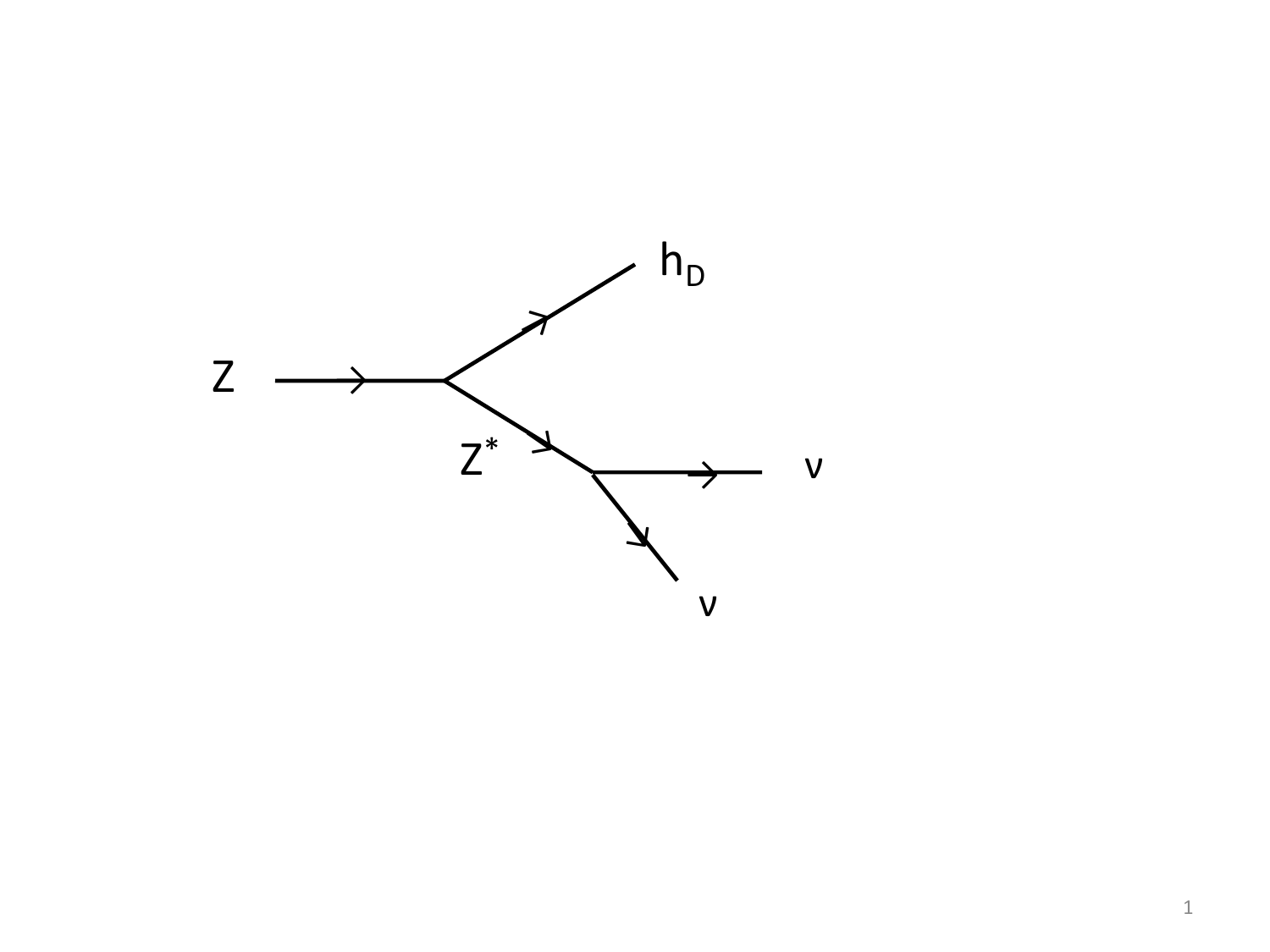}}
\vspace*{-3.5cm}
\caption{Diagram leading to a 3-body invisible $Z$ decay arising from SM and dark Higgs mixing with the off-shell SM $Z$ decaying to the familiar 3 generations of neutrinos.  }
\label{fig1}
\end{figure}

We begin with a rather simple case:  as well-known, the SM Higgs doublet, $H$, and (for simplicity here considered to be just a single field) the complex, SM singlet dark Higgs, $S$, fields can 
mix due to their non-zero vevs, $v,v_s$, respectively, through the standard quartic term appearing in the scalar potential, \ie,  
\begin{equation}
 U_{HS} = ~\lambda_{HS} H^\dagger H S^\dagger S \,,
\end{equation}
where we imagine that $S= (v_s+h_D+iG_V)/\sqrt 2$, with $h_D$ being the rather light, $\lsim 1$ GeV, physical dark Higgs while $G_V\sim V_L$ is essentially the Goldstone boson eaten by the DP to 
obtain its mass in the limit that $v_s/v <<1$.  Upon diagonalization of the Higgs mass matrix, such a term then leads to a coupling between the SM $Z$ and the dark Higgs, \ie, $ZZh_D$, and thus we 
can imagine a 3-body decay of the form $Z\to h_DZ^*, Z^*\to \bar \nu \nu$, as is shown in Fig.~\ref{fig1}. This is just the dark Higgs analog of the Bjorken process\cite{bj}, proposed 50 years ago as a 
channel for light SM Higgs searches. Since $h_D$ decays to pairs of dark photons which, in turn, are assumed to decay to DM, this final state is clearly invisible. Since all the particles in the final 
state are essentially massless, as $m_{h_D,V}^2/m_Z^2 <<1$, it is easy to estimate the size of the branching fraction for this decay mode\cite{bj}; summing over the three generations of neutrinos we 
find that 
\begin{equation}
B_{inv}=B(Z\to h_D \bar \nu \nu) \simeq 2 \cdot 10^{-3} ~\theta_{hh_D}^2 \,,
\end{equation}
with $\theta_{hh_D}$ being the mixing angle between the SM and dark Higgs. As is well-known, however, this angle must be quite small to avoid inducing a large invisible branching fraction, $B_h$, 
for the SM Higgs for the decays into the DP and/or dark Higgs pairs final states. ATLAS and CMS measurements\cite{Invisible} inform us that, roughly, $B_h \lsim 0.1$ and thus we obtain 
the bound that $|\theta_{hh_D}|\lsim 5\cdot 10^{-4} ~(\frac{v_s/v}{10^{-2}})~ \sqrt{B_h/0.1}$ from which we can then determine that the new contribution to the invisible width of the $Z$ is given by 
$\Delta \Gamma_{inv}\lsim 1.3\cdot 10^{-3}$ keV, which is very far below the potential $\simeq 25 $ keV sensitivity of Tera-Z as discussed above. Thus we conclude that this particular process is 
not relevant as a new source of invisible $Z$ decays.

\subsection{$Z$ Mixing with the Dark Photon}

A somewhat different but also familiar process, the mass and/or KM of the SM $Z$ with the $U(1)_D$ DP, $V$, can also potentially lead to invisible $Z$ decays since, as we will assume here, the DP 
dominantly decays to DM. Generically, suppressing Lorentz indices, before spontaneous symmetry breaking the kinetic terms for the neutral SM gauge fields plus the DP can be written as 
\begin{equation}
{\cal L}_{KM}=-\frac{1}{4}\hat B \hat B -\frac{1}{4}\hat W_3 \hat W_3 -\frac{1}{4}\hat V \hat V +\frac{\alpha}{2}\hat B \hat V +\frac{\beta}{2}\hat W_3 \hat V\,,
\end{equation}
where the parameters $|\alpha,\beta| \lsim 10^{-3}$ in the present model describe the results of vacuum polarization-like (potentially finite) loop-induced KM produced by heavy scalar and/or 
vector-like fermion PM fields having non-zero values of $T_3$ and/or electric charge, $Q$, plus a non-zero $Q_D$, the dark charge to which the DP, $V$, couples.  In terms of the quantities below, we 
can generically express the parameters $\alpha,\beta$ as quantum number weighted sums over the PM fields, $P$, appearing in the relevant loops with $G$ being the appropriate loop function 
for the contributing fermions or complex scalars:
\begin{equation}
\alpha=g_Dgt_w\sum_P ~Q_D(P)\Big(Q(P)-T_3(P)\Big)~G(P),~~~\beta=g_Dg\sum_P~Q_D(P)T_3(P)~G(P)\,.
\end{equation}
with $g[g_D]$ being the $SU(2)[U(1)_D]$ gauge coupling.  As is well-known, since both $\alpha,\beta$ are loop suppressed, \ie, quite small, 
this KM can be removed to leading order by a simple linear set of field redefinitions: $\hat B \to B+\alpha V$, $\hat W_3 \to W_3+\beta V$, $\hat V \to V_0$. Recalling that in the SM $Q=T_3+Y/2$, 
$g_Y=g\tan \theta_w=gt_w$ being the $U(1)_Y$ hypercharge gauge coupling expressed in terms of the corresponding $SU(2)_L$ gauge coupling, $g$. With $\theta_w$ being the 
usual weak mixing angle, $s_w=\sin \theta_w$, 
\etc, so that $e=gs_w$ as usual and employing the standard definitions, $W_3=Z_0c_w+As_w$, $B=ac_w-Z_0s_w$, with $A$ being the familiar photon, we can schematically write the 
relevant interaction piece of the covariant derivative as (again dropping Lorentz indices) 
\begin{equation}
{\cal D}_{int}=eQA+\frac{g}{c_w}(T_3-s_w^2Q)Z_0+\Big[g_DQ_D+\frac{g}{c_w}\big(\alpha s_wQ+(\beta c_w-\alpha s_w)T_3\big)\Big]V_0\,.
\end{equation}

Next, we imagine that the Higgs sector for this setup consists of the usual SM weak isodoublet having a $T_3=1/2$ vev, $v$, with $Q_D=0$, a dark Higgs weak isosinglet decomposing as, 
$H_s=(v_s+h_s+ia_s)/\sqrt 2$, 
having $Q_D=Q_D^s\neq 0$ and now also a second dark Higgs isodoublet, $H_d$, whose neutral $T_3=1/2$ component similarly decomposes as $H_d^0=(v_d+h_d+ia_d)/\sqrt 2$, 
and which also carries a dark charge $Q_D^d$. Note that while $h_{d,s}$ are real, CP-even scalars, $a_{s,d}$ are CP-odd. In the limit that $v_{d,s}<<v$, via the quartic terms in the scalar potential, 
one linear combination of $h_{d,s}$ becomes the light dark Higgs while the other, $H$, can generally be several times heavier than the SM Higgs, $h$. Similarly, in this same limit, one linear 
combination of CP-odd fields, $a_{d,s}$, becomes the Goldstone boson for the DP, $G_V$, while the second, $A$, generally obtains a large mass quite similar to that of the new heavy neutral scalar. 
This Higgs sector is completely analogous to that employed by Davoudiasl and Marciano\cite{hoomie} early on and is sufficiently general to make the 
initial arguments that we wish to below. Note that whereas the vev $v_s$ only contributes to the DP mass and $v$ only to the $Z$ mass, $v_d$ contributes to both the DP and $Z$ masses and 
so induces $Z_0-V_0$ mass mixing. Decoupling the usual SM massless photon, $A$, the remaining $2\times 2$ mass-squared matrix in the $Z_0-V_0$ basis is then given by
\begin{equation}
{\cal M}^2=\frac{g^2v^2}{4c_w^2}~\begin{pmatrix} 1+D& rD+(c_w \beta-s_w \alpha)\\ rD+(c_w \beta-s_w \alpha) & \gamma^2 \\ \end{pmatrix}\,,
\end{equation}
where we have defined the ratios $D(S)=v_{d(s)}^2/v^2\sim 10^{-5}-10^{-4}<<1$ as we expect that $v_{d,s}\lsim 1$ GeV, $r=2c_wQ_D^dg_D/g$, which we expect to be $O(1)$, and thus we have that 
$\gamma^2=r^2[D+SQ_D^{s~2}/Q_D^{d~2}]$. 

To a very good approximation, this matrix can be diagonalized via a small rotation, $Z_0=Z-\chi V, ~V_0=V+\chi Z$, with $\chi=rD+(c_w \beta -s_w \alpha)$ just being the off-diagonal element 
of ${\cal M}^2$ above, so that, to leading order in the small parameters, the SM $Z$ mass remains unaltered and the DP mass-squared is just given by its unmixed value, \ie, 
$m_V^2=g_D^2[(Q_D^d v_d)^2+(Q_D^s v_s)^2]$. For simplicity we will take $Q_D^d/Q_D^s=\lambda =\pm 1$ in the discussion that follows. 
In this same small parameter expansion limit, the $Z$ and $V$ mass eigenstates are now found to couple to the combinations 
\begin{equation}
\Big[\frac{g}{c_w}\big(T_3-s_w^2Q\big)+\chi g_DQ_D\Big]Z +\Big[g_DQ_D+e(s_w\beta+c_w\alpha)Q+\frac{g}{c_w}\big(T_3-s_w^2Q\big)(-rD)\Big]V\,,
\end{equation}
so that we can identify the usual KM parameter for the photon as $\epsilon=\epsilon_\gamma=s_w\beta+c_w\alpha \simeq 10^{-4}-10^{-3}$ while the corresponding KM parameter for the $Z$ is 
just $\epsilon_Z=c_w\beta-s_w\alpha$, and generally having a similar magnitude.  Since the $Z$ can now couple to the dark sector SM singlet fields with $Q_D\neq 0$ this opens up a new window for 
invisible decays, \eg, $Z$ decay to DM. However, such decays are also suppressed by an overall factor of $\chi^2\lsim 10^{-7}$ or so leading to partial widths which are then far below the $\simeq 25$ 
keV scale and are so are undetectable even at Tera-Z. Thus we conclude the small $Z-V$ mass/kinetic mixing expected in this general setup cannot lead to a significantly large contribution to the 
invisible width of the $Z$.

\subsection{An Aside: A Different Decay Mode}

There is, however, another source for invisible $Z$ decays, to a {\it different} final state, hiding in the previous discussion which is due to the dark Higgs doublet, $H_d$, whose $T_3=1/2$ 
neutral member, $H_d^0$, obtained 
the vev $v_d$ employed above, as was discussed indirectly in our earlier work\cite{Rueter:2020qhf}.  As we discussed above, in the limit that $v_{d,s}<<v$, a linear combination of the scalars 
$h_{d,s}$ becomes the dark Higgs, $h_D$, while a linear combination of $a_{d,s}$ becomes the Goldstone boson for the DP, $G_V\sim V_L$ due to the Equivalence Theorem\cite{GBET}, 
both of which are far lighter than the $Z$. Since both $h_D$ and 
$G_V$ contain some admixture of the two $T_3=1/2$ states, the tree-level decay $Z\to h_DG_V$ becomes possible and, further, since both $h_D$ and $G_V$ (\ie, $V$) decay invisibly, this mode 
will contribute to the invisible width of the $Z$. It is trivial to see that this partial width is potentially quite large and will easily conflict with current experimental bounds; in the massless, {\it unmixed} 
case where, \eg, $H_s$ is {\it absent} or is vev-less, clearly we can the identify $h_D=h_d$ and $G_V=a_d$, and one then observes that 
\begin{equation}
\Delta \Gamma(Z\to inv)=\Gamma(Z\to h_da_d)=\Gamma(Z\to h_DG_V)=\frac{1}{2} \Gamma(Z\to \bar \nu \nu)\simeq 83.3~[2T_3(h_d)]^2~{\rm MeV}\,,
\end{equation}
where in the present case $T_3(h_d)=1/2$ and which we see (very) far exceeds the current $95\%$ CL upper bound of $\simeq 2.4$ MeV\cite{limit}. This means that $h_D$ and $G_V$ must likely 
be {\it mostly} composed of the SM singlets $h_s$ and $a_s$ in the model above in order to suppress this decay rate. Clearly, in the opposite limit where $h_D=h_s$ and $G_V=a_s$ this partial 
width completely vanishes as these fields have $T_3=0$, being isosinglets, and so will no longer couple to the $Z$ at tree level. 

What happens in the more realistic mixed case? In the model example above with one dark doublet and one dark singlet from the last Subsection, we note that the detailed structure of an 
important term in the scalar potential that leads to mixing is dependent upon the choice $\lambda=\pm 1$, \ie, whether $H_d$ and $H_s$ have the same or opposite dark charges. With 
$\sigma_0>0$ being a dimensionless parameter this term is
\begin{equation}
U_{mix}=\frac{-\sigma_0 v}{\sqrt 2} \Big( H^\dagger H_dH_s^\dagger+h.c.\Big)~~ {\rm for} ~\lambda=1, ~~=\frac{-\sigma v}{\sqrt 2} \Big( H^\dagger H_dH_s+h.c.\Big)~~ {\rm for}~ \lambda=-1\,.
\end{equation}
In either case, writing $h_d=Hc_h-h_Ds_h, ~h_s=h_Dc_h+Hs_h$ and analogously, $a_d=Ac_a-G_Vs_a,~a_s=G_Vc_a+As_a$ (with $H,A$ as described above), one finds to leading order in the 
small ratios $v_{s,d}/v <<1$ that indeed $G_V$ is massless as expected, that there are two essentially degenerate heavy states, 
\begin{equation}
m_{H,A}^2\simeq \sigma_0v^2\Big(t+1/t\Big)\,,
\end{equation}
and the value of  $m_{h_D}$ is set by the $v_{d,s}$ mass scale as it should, \ie, 
\begin{equation}
m_{h_D}^2\simeq \frac{2\lambda_dv_d^2t+2\lambda_sv_s^2/t+2\kappa v_dv_s}{(t+1/t)}\,,
\end{equation}
with the parameters $\lambda_{d,s}, \kappa$ all originating from the relevant the quartic pieces of the scalar potential:
\begin{equation}
U_{quartic}=\lambda_d(H_d^\dagger H_d)^2+\lambda_s(H_s^\dagger H_s)^2+\kappa (H_d^\dagger H_d) (H_s^\dagger H_s)+....\,,
\end{equation}
together with the identification of the two mixing angles as $t_h=-t$, $t_a=-\lambda t$ with $t=v_d/v_s$. These mixings result in a suppression of the 
$Z\to h_DG_V$ partial width by a factor of $s_h^2s_a^2=t^4/(1+t^2)^2$ (since $\lambda^2=1$) so that to satisfy the current limit we must have $t=v_d/v_s < 0.45$ or so, also further implying 
that, \eg,  $Z-V$ mass mixing is likely significantly suppressed relative to that arising from KM. If nothing new is observed within the expected sensitivity  at Tera-Z this bound will improve to 
roughly $t <  0.13$, further tightening the requirement 
that the dark Higgs be mostly isosinglet.  We note again that this decay into invisible final states {\it cannot} happen at tree level when the dark Higgs is 
either a pure isosinglet or is a linear combination of nontrivial multiplet members all of which have $T_3=0$ as in, \eg,  the model employed in Refs.\cite{Rizzo:2022jti,Rizzo:2026dsj}. 
One possible caveat to this is when we have two (or more) non-trivial dark Higgs representations, for example, 2 dark isodoublets with opposite values of $Q_D$ as in the model employed in 
Ref.\cite{Rueter:2020qhf}. There it was shown that the $Z\to h_DG_V$ partial width is instead suppressed by a factor of $(t^2-1)^2/(t^2+1)^2$ where now $t$ is again just the ratio of the two dark 
doublet vevs.  As found there, adjusting this vev ratio ever close to unity will allow one to escape the present and potential future experimental constraints on this decay but this can be seen 
to come with a significant fine-tuning cost.

\section{$Z$ Couplings to the Dark Sector via PM Loops}

In this Section we will consider how PM fields can generate a small contribution to the invisible width of the $Z$, again via the, now loop-induced, $h_DG_V$ final state, whose magnitude satisfies 
existing constraints but will be potentially experimentally accessible at Tera-Z. Unlike many of the tree level mixing scenarios, these loop effects directly probe the nature of the PM fields themselves. 
To be specific, in this Section, we consider an abbreviated sub-class of the `$E_6$-inspired' scenario\cite{Rueter:2019wdf,Rizzo:2026dsj} as was frequently employed in our earlier work. 
Here the PM fields will consist of, in $SU(5)$ language, a ${\bf 5}+\bar {\bf 5}$ pair of vector-like fermions, analogous to the usual SM $\bar {\bf 5}$ (shown below for later use), which are assumed to 
lie at the mass scale of $\sim$ a few TeV, resulting from the breaking of a larger dark gauge group\cite{Rueter:2019wdf}, $G_D$ down to $U(1)_D$:
\begin{equation}
{\bf 5}=\begin{pmatrix} D\\N^c \\ E^c\\\end{pmatrix}_L, ~\bar {\bf 5} =\begin{pmatrix}  D^c\\N \\ E\\\end{pmatrix}_L, ~~{\bar {\bf 5}}_{SM}=\begin{pmatrix} d^c\\\nu\\ e \\\end{pmatrix}_L\,,
\end{equation}
but where here, for simplicity, the dark gauge sector consists of just the familiar $U(1)_D$. We will sometimes employ the familiar left-handed field notation for these fermions and, frequently, we will 
denote the $SU(2)$ isodoublets as $L=(N,E)^T$ and $\ell=(\nu,e)^T$. Note that we take $Q_D(D)=-Q_D(L)=1$ from the `$E_6$-like' discussion without loss of generality while, as 
usual, $Q_D(SM)=0$.  This setup is also seen to lead to a finite value for $\epsilon_\gamma$ in the experimentally interesting range for comparable values of the $L$ and $D$ masses. 
In general, there are $N_g=3$ generations of such PM fields (which we'll return to below), but for simplicity we merely describe this setup for the case of a single generation, thus ignoring any 
potential mixing effects; however, for any numerical estimates below we will assume that these three PM fermion generations of a given type, $L_{i=1-3},D_{i=1-3}$, are present and are all 
degenerate in mass.  The PM fields are accompanied by the dark Higgs field, $S$, which is an $SU(5)$, hence SM, singlet, having $Q_D=-1$. From the previous Section, we know that if 
$h_D,G_V$ arise purely from $SU(2)_L$ singlets then there will be no tree-level decay of the form $Z\to h_DG_V$ as the relevant coupling is absent. Here, $S$ plays two essential roles in this setup: 
first, its small vev, $v_s\lsim 1$ GeV, will as usual break $U(1)_D$ generating the DP mass, $m_V=g_Dv_s$; second, this same vev generates a small mixing between the PM fields and those 
of the SM with the same SM quantum numbers, \ie, color, weak isospin and electric charge, allowing the heavier PM fields to decay. After $SU(5)$ breaking and allowing (at this level of discussion) for
 bare TeV-scale mass terms for the PM fields the relevant couplings are just (again in left-handed field notation) given by 
\begin{equation}
{\cal L}_{mass}= m_D D^cD_L+m_L L^cL_L+(y_Dd^cD_L+y_L L^c \ell_L)S +h.c.\,,
\end{equation}
such that $N,E$ are degenerate in mass before mixing, \ie, $m_N=m_E=m_L$ (and which remains true to a very high accuracy even after mixing) and $y_{D,L}$ are assumed to be two, $O(1)$, 
but generally distinct, Yukawa couplings allowing the PM to decay promptly. Again, when we consider the 
case of 3 generations, for simplicity we will assume that these couplings are all flavor-diagonal and identical, \ie, $y_{L_{i=1-3}}=y_L$, \etc. Since only the $Q=-1$ and $Q= -1/3$ SM 
fermions will be of relevance in the discussion below and their squared masses are all far below that of the $Z$, it will be sufficient for our purposes to give them small common masses,  
$m_{d,\ell}$, when needed in the analysis below. Numerically, we will generically work to leading order in the limits $r_{D,L}=m_Z^2/m_{L,D}^2\lsim 10^{-2}<<1$ and $m_{d,\ell}^2/m_Z^2<<1$ 
in the discussion that follows.

\begin{figure}[htbp] 
\hspace*{0.5cm}\centerline{\includegraphics[width=4.5in,angle=0]{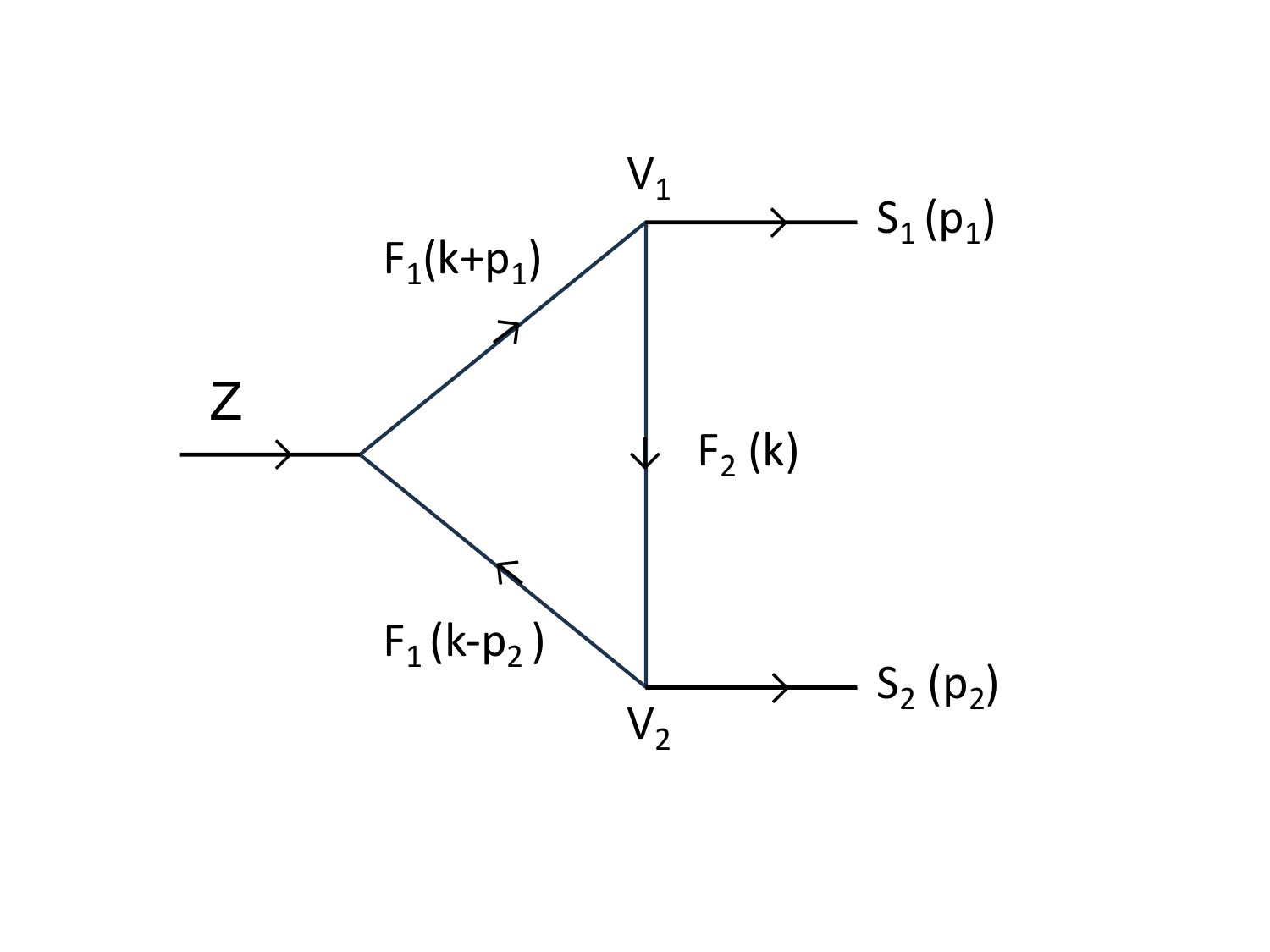}}
\vspace*{-2.0cm}
\caption{Generic PM-induced triangle loop diagram leading to the decay $Z\to SS^\dagger \to {\rm invisible}$. The possible identity of the fermions $F_{1,2}$ in the loop, the final state 
scalars, $S_{1,2}$, and the helicities occuring at the two vertices, $V_{1,2}$, are given in Table~\ref{loopstab}. The arrows label the momentum flow in the graph. }
\label{fig2}
\end{figure}

Recalling that in the $v_s<<v$ limit, apart from the vev itself, $S\simeq (h_D+iG_V)/\sqrt 2$ so that the weak scale we can think of $Z\to h_DG_V$ as just the process $Z\to S^\dagger S$ and 
neglect the squared masses 
associated with either $h_D$ or $G_V=V_L$ relative to that of the $Z$. Further, since $E,N$ are essentially degenerate to an excellent approximation we can treat them in combination as just the 
single entity, $L$, \ie, we simple add their contributions and attributed them to $L$. In a likewise fashion, since $m_\nu,m_e <<m_Z$, we can also treat them in combination as the single entity 
$\ell$, which we obtain by simply adding those from the individual fermions $\nu$ and $e$. This 
being the case, there are then 4 distinct, one-loop triangle graphs that will contribute to this process that take the generic form shown in Fig.~\ref{fig2} with the identity of the various $F_i,S_i$ as 
given in Table~\ref{loopstab}. Due to the cancellations between the SM and PM contributions, the sum of these amplitudes is found to yield a finite result. 
The  values of the $V_i$ tell us the helicity projections, $L$ or $R$,  associated with the vertices where either $S$ of $S^\dagger$ are emitted from the 
fermion line.  Of course, the SM $Z\bar F_1F_1$ vertex has, in general, both helicities contributing whether or not $F_1$ is a SM fermion (in which case the left- and right-handed couplings are distinct) 
or a PM fermion (in which case both of these couplings are the same).  If the resulting $ZS^\dagger S$ vertex can be parameterized as 
\begin{equation}
i\Big(\frac{g}{c_w}\Big)K~\big(p_S-p_{S^\dagger}\big)_\mu \cdot \epsilon(Z)^\mu\,,
\end{equation}
with $K$ being a generally complex quantity arising from the diagram sum, $p_i$'s being the final state particle momenta and $\epsilon(Z)^\mu$ the familiar $Z$'s polarization vector,
then the new loop induced contribution to the invisible width of the $Z$ (again taking $m_{h_D,V}^2<<m_Z^2$) is, with this normalization, given numerically by
\begin{equation}
\Gamma(Z\to S^\dagger S)=2|K|^2 \Gamma_\nu\ \simeq 333.3 |K|^2 ~{\rm MeV}\,,
\end{equation}
and thus our goal is to determine the typical range of values for $|K|$ in this setup; note that Tera-Z will be sensitive down to values of roughly $|K| \simeq 0.87\cdot 10^{-2}$ corresponding a $25$ keV 
partial width.

\begin{table}
\begin{center}
\caption{Possible identities of the fermions, scalars and vertex helicities in the loop appearing in Fig.~\ref{fig2}.}
\label{loopstab}
\begin{tabular}{ c c c c c c c }
\hline
 & $F_1$ & $F_2$ & $S_1$ & $S_2$ & $V_1$ & $V_2$\vspace{0.1cm} \\
\hline
 i)& $D$ & $d$ & $S^\dagger$ & $S$ & R & L\vspace{0.1cm}\\
 ii)& $d$ & $D$ & $ S $ & $S^\dagger$ & L & R\vspace{0.1cm}\\
 iii)& $L$ & $\ell$ & $S$ & $S^\dagger$ & L & R\vspace{0.1cm}\\
 iv)& $\ell$ & $L$ & $S^\dagger$ & S & R & L\vspace{0.1cm}\\
\hline 
\end{tabular}
\end{center}
\end{table}   

Taking the limit $m_{d,\ell}^2<<m_Z^2$, these contributions will generally fall into two classes depending upon whether or not $F_1$ is a heavy PM field, thus appearing on two of the internal lines; 
if so, then both of the chiral couplings to the $Z$, $c_{L,R}(D,L)$ (which happen to be identical in this case), will contribute to the amplitude whereas,  if not, then only $c_R(d)$ and $c_L(\ell)$ 
will appear. Combining the $N,E$ and $\nu,e$ contributions as described above, \eg, $c_L(\nu,N)+c_L(e,E)=c_L(\ell,L)$, \etc, all of the relevant chiral couplings are found to have the same value 
(once the scaling by the appropriate $N_c=1,3$ color factors is included) and are equal to $x_w=s_w^2$. As an example, in the $F_1=D$ case, employing the labeling as in Fig.~\ref{fig2}, 
the amplitude is found to proportional to the expression
\begin{equation}
\sim  ~{{Tr~\gamma_\mu \Big[(\kslash-\pslash_2)+m_D\Big]P_R(\kslash+m_d)P_L\Big[(\kslash+\pslash_1)+m_D\Big]}\over{(k^2+2k\cdot Q_1-M_1^2)^3}} \,,
\end{equation}
where $k$ is the usual loop momentum to be integrated over, $Q_1=-xp_2+yp_1$ with $Q_1^2=-xym_Z^2$ and $M_1^2=(x+y)m_D^2+(1-x-y)m_d^2$, with $x,y$ being the standard 
Feynman parameters. It is to be noted that here $p_1=p_{S^\dagger}$ while $p_2=p_S$; here and in what follows we will set $p_1^2=p_2^2=0$. 
Similarly, when, \eg, $F_1=d$, we find the corresponding amplitude is proportional to 
\begin{equation}
\sim  ~{{Tr~\gamma_\mu \Big[(\kslash-\pslash_2)+m_d\Big]P_L(\kslash+m_D)P_R\Big[(\kslash+\pslash_1)+m_d\Big]}\over{(k^2+2k\cdot Q_2-M_2^2)^3}} \,,
\end{equation}
with the roles of $p_{1,2}$ being interchanged, \ie, $S\leftrightarrow S^\dagger$, and where now also $Q_2=Q_1(p_S\leftrightarrow p_{S^\dagger})$,  with $Q_2^2=-xym_Z^2$ and 
$M_2^2=(1-x-y)m_D^2+(x+y)m_d^2$ as might be expected.

After integration over the loop momentum and now including the overall numerical factors, the relevant, now finite, piece of the sum of the two $F_1=D,d$ contributions to the amplitude can be 
written as  
\begin{equation}
iN_g~\Big(\frac{g}{c_w}\Big)~\Bigg(\frac{s_w^2 ~y_D^2}{16\pi^2}\Bigg)~(I_{1D}+I_{2D})~\Delta_\mu\,,
\end{equation}
where $\Delta=p_1-p_2$ is the 4-momentum difference of the particles in the final state as seen in Fig~\ref{fig2} and $I_{1D},I_{2D}$ are just the familiar looking parametric integrals
\begin{equation}
I_{1D}=\int_0^1 dx \int_0^{1-x} dy ~ (4-3x-3y) ~~{\rm log}\Bigg[ \frac{x+y-r_Dxy}{1-x-y -r_Dxy}\Bigg], ~~I_{2D}=-\int_0^1 dx \int_0^{1-x} dy ~~\frac{x+y}{x+y-r_Dxy}\,.
\end{equation}
Note that when, \eg, $D(d)\to L(\ell)$, the helicities at the $V_{1,2}$ vertices are interchanged and even more importantly $S\leftrightarrow S^\dagger$ leading to an overall change of sign due to 
the outgoing momentum assignments, \ie, $p_1=p_S \to p_1=p_{S^\dagger}$, \etc. This implies that we should expect our result to vanish in the $SU(5)$ limit when $m_D=m_L$ and $y_D=y_L$; this 
expectation is in fact realized in the final expression obtained below.

Indeed, now summing over these four contributions, $F_1=(D,d,L,\ell)$, 
performing the necessary integrations, then expanding to leading order in the large mass ratios, $r_{D,L}^{-1}$, and dropping all power suppressed terms, we find that we can express $K$ as just 
\begin{equation}
K\simeq \frac{N_gs_w^2}{16\pi}~\Bigg[\frac{1}{\pi}\Big(y_D^2(\log r_D-17/6)-y_L^2(\log r_L-17/6)\Big) +i(y_D^2-y_L^2)\Bigg]\,, 
\end{equation}
which we see vanishes in the $SU(5)$ limit as expected. Of course, as is well-known, even if taken serious at high scales, any $SU(5)$ symmetery is very badly broken at the low energies 
$\sim $ a few TeV or below that we care about here by expected $O(1)$ factors. 

As a simple numerical example (with $N_g=3$), taking the suggestive values of $m_D=2$ TeV and $m_L=1.5$ TeV to safely satisfy the existing experimental LHC search constraints on these 
PM states yet not be too far away in mass for future direct searches, as summarized in, \eg, Ref.\cite{Rizzo:2022qan} for these types of fermions, together with the O(1) 
values $y_D=1$ and $y_L=0.5$ of the Yukawa couplings for demonstration purposes, one finds that 
\begin{equation}
K\simeq 0.0138\big(-2.20+0.75i\big) ~~\to~~ |K|\simeq 0.032\,.
\end{equation}
From this we obtain $\Delta \Gamma(Z\to ~inv)\simeq 343$~ keV, far exceeding the Tera-Z sensitivity limit yet still quite safe with respect to the current search constraints. As might be expected, 
varying these numerical assumptions by factors of order unity will not change this result in a more than semi-quantitative manner (away from the $SU(5)$ symmetric limits, $y_D\simeq y_L$ and 
$m_D\simeq m_L$) yielding predictions that will still generally lie within this experimentally accessible (and allowed) range, \ie, $0.87\cdot 10^{-2} \leq |K| \leq 8.5\cdot 10^{-2}$. Interestingly, 
these same input values will lead to a finite and calculable KM parameter value of $\epsilon_\gamma \simeq (10^{-4}-10^{-3})~(g_D/e)$ which also lies in the experimentally relevant range. 

\begin{figure}[htbp]
\centerline{\includegraphics[width=5.0in,angle=0]{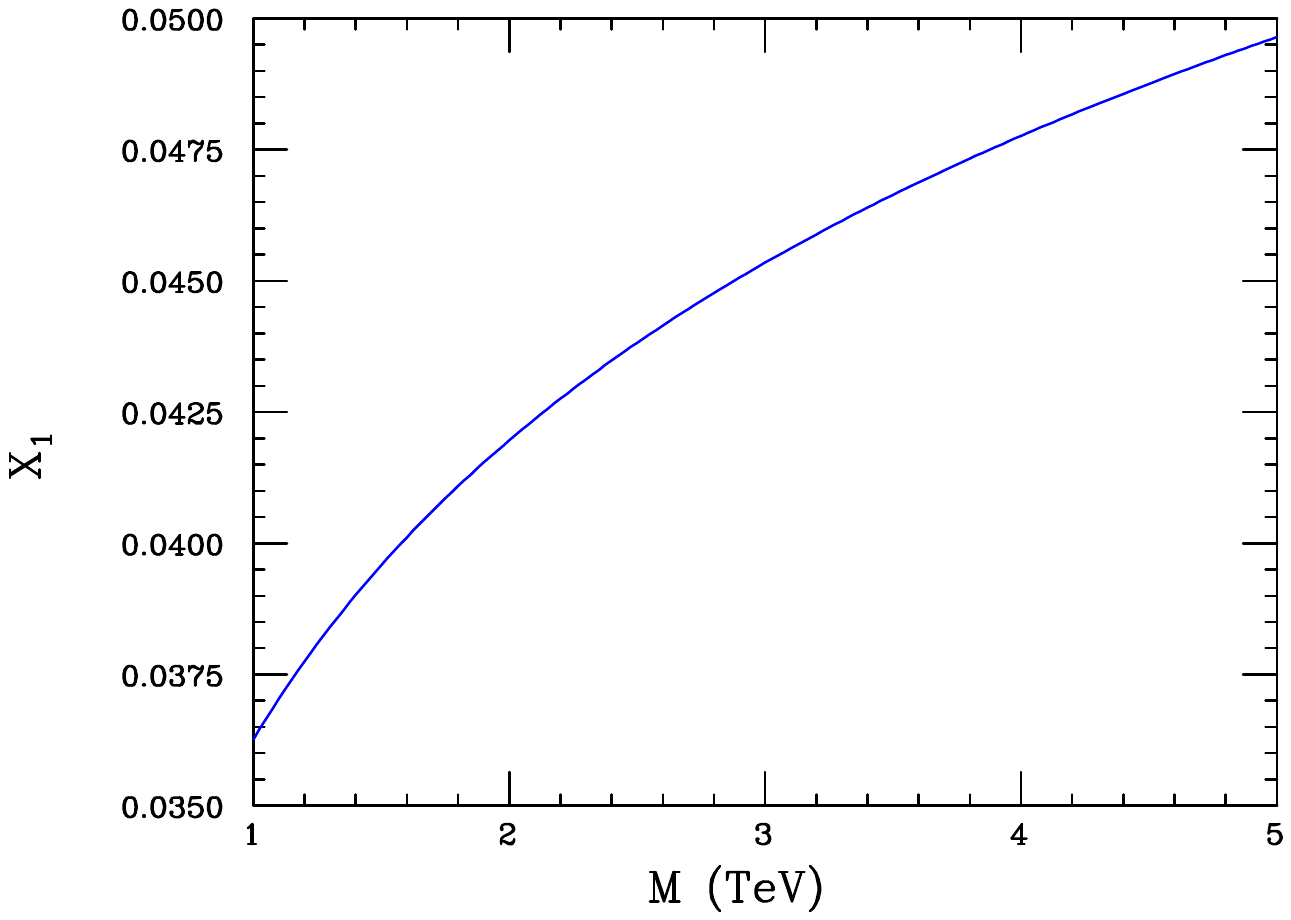}}
\vspace*{-1.5cm}
\centerline{\includegraphics[width=5.0in,angle=0]{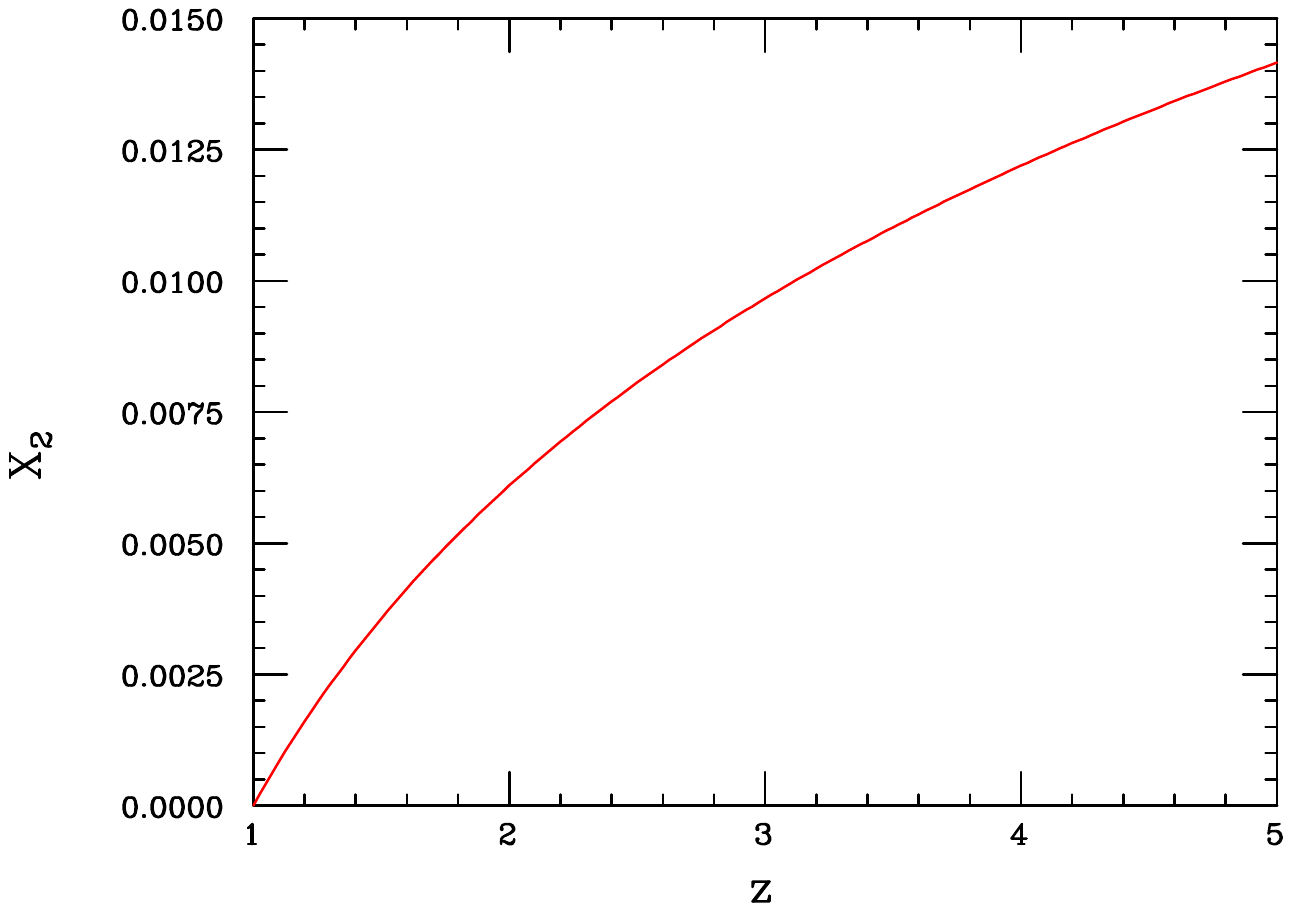}}
\vspace*{-1.3cm}
\caption{(Top) The quantity $X_1$, as defined in the text, as a function of the common PM mass, $M=m_D=m_L$. (Bottom) The quantity $X_2$, as defined in the text, as a function of the ratio 
of the $D$ and $L$ PM masses, $z \geq 1$.}
\label{invis}
\end{figure}

To get a better idea of the potential range of predicted $|K|$ values we contemplate two simple, yet essentially `opposite extreme', examples:  First, we consider the case where $m_D=m_L=M$ while 
$y_D \neq y_L$ and construct the quantity $X_1=|K|/|(y_D^2-y_L^2)|$ which then only depends upon the single parameter $M$. This quantity is shown in the upper panel of Fig.~\ref{invis} where 
we see that, for values of $M$ of order a few TeV or so,  roughly extends over $X_1=0.035 - 0.050$ so that a significant range of the Yukawa coupling-squared difference would yield shifts in the 
invisible branching fractions which would be accessible to Tera-Z observation while also satisfying present experimental bounds.  In the second example, we consider the opposite limiting case 
where $y_D=y_L=y$ while allowing $m_D\neq m_L$ and correspondingly define the quantity $X_2=|K|/y^2$ which is then a function only of the dimensionless ratio of the masses $m_D$ and $m_L$, 
here called $z$. This quantity $X_2$ as a function of $z$ is then shown in the lower panel of Fig.~\ref{invis} where we see that for $z\leq 5$ it ranges from zero (corresponding to the exact 
$SU(5)$ limit) to roughly $\simeq 0.014$. Clearly, for smaller values of $z$, a large value of the common Yukawa coupling, $y$, would be required to push the value of $|K|$ into the range needed 
to make this new contribution to the invisible width of the $Z$ observable at Tera-Z; for example, when $z=1.5$, $y\gsim 1.6$ or so would be required. For larger $z$, more modest values of $y$ 
would then only be needed. It is easily seen that even larger values of $z$ would also be allowed for appropriate ranges of $y$.  

Interestingly, this same simple model can lead to a number of additional alterations to $Z$ decays due to the same set of particles and interactions introduced above; these will be discussed 
elsewhere\cite{next}. 

Certainly this example toy PM model is not unique in yielding both a calculable value for the strength of the KM as well as finite results for this new contribution to invisible partial width of the $Z$ in this 
interesting range via loops and we expect that other more complex and realistic setups will produce predictions which will also overlap these experimentally interesting ranges.

\section{Discussion and Conclusion}

The kinetic mixing of the SM photon with a light dark photon provides an interesting portal which allows for the existence of sub-GeV thermal dark matter. Embedding such a setup into more 
UV-complete models can yield a large range of possible new signatures related to the portal matter fields generating this loop-order kinetic mixing. In addition to the direct production of numerous 
new heavy states at high energy colliders (\eg, the PM itself, the new gauge bosons associated with the enlarged dark group $G_D$ beyond the DP, and the associated new heavy Higgs fields 
needed for breaking this symmetry), new, other more indirect signatures of such a general setup are also to be anticipated - more likely than not from additional loop effects of PM. Some of 
these possibilities we have already encountered in earlier analyses such as light Dirac masses for neutrinos\cite{Rizzo:2026mza} and new, non-KM interactions between the SM and dark sector 
fields via anomalous dark moments\cite{Rizzo:2021lob,Rizzo:2026dsj}. While other such possibilities are currently being examined\cite{next}, here we have considered the potential PM setup 
contribution to the invisible width of the $Z$, $\Delta \Gamma_{inv}$. It is anticipated that precision measurements made at the FCC-ee, running in the Tera-Z mode, will be able to extend the 
current reach for such effects by roughly two orders of magnitude beyond present sensitivities which will open a new window to search for light dark sector physics.  
 
 In this paper, we have considered two general frameworks for generating such KM-PM model contributions to $\Delta \Gamma_{inv}$: mixing- and loop-induced effects. The simplest and most 
 familiar possibility, mixing the SM Higgs with an isosinglet dark Higgs through quartic terms in the scalar potential was shown to 
 lead to a contribution to $\Delta \Gamma_{inv}$ which was several orders of magnitude too small due to the constraint on the invisible width of the SM Higgs from LHC measurements. The second 
 scenario, mass and/or KM between the DP and the SM $Z$, thus inducing a $Z$ coupling directly to light DM, was also found to be too small due to the required size of the DP-SM KM as well as 
 the small ratio of the 
 DP to SM $Z$ squared masses, again by several orders of magnitude. As part of this analysis it was shown that a potentially very large contribution to $\Delta \Gamma_{inv}$ is possible when 
 dark Higgs fields are present which transform non-trivially under the SM $SU(2)_L$ gauge group whose $T_3 \neq 0$ components obtain vevs that (also) break $U(1)_D$ and have a sizable 
 overlap with the light dark Higgs mass eigenstate. In general in such cases, such contributions to $\Delta \Gamma_{inv}$ are, in fact. found to be {\it too} large and will violate current bounds 
 unless the parameters of the model are somewhat fine tuned.  As an example of potential loop-induced effects, we considered a simplified toy model version of the `$E_6$-inspired' scenario 
 employed in earlier work where the three generations of PM fermion fields each lie in ${\bf 5}+\bar {\bf 5}$ pairs of $SU(5)$. For typical parameter choices this setup was shown to yield a finite 
 and calculable value for 
 $\Delta \Gamma_{inv}$ in the range safely below the current bound while also within the range of sensitivity accessible to Tera-Z. In addition, the calculable KM parameter for this setup is also 
 found to be finite and in an experimentally interesting range for light DM and DP.  This is, of course, only just one simple toy example of a larger class of possible models that have these same 
 semi-quantitative predictions deserving further attention elsewhere.

Hopefully, direct or indirect signatures of the dark sector will soon make themselves known to us.

\section*{Acknowledgements}
The author would like to particularly thank J.L. Hewett for both hospitality and wide ranging discussions. This work was supported by the Department of Energy, Contract DE-AC02-76SF00515.




\begin{thebibliography}{99}



\bibitem{Planck:2018vyg}
N.~Aghanim \textit{et al.} [Planck],
Astron. Astrophys. \textbf{641}, A6 (2020)
[erratum: Astron. Astrophys. \textbf{652}, C4 (2021)]
[arXiv:1807.06209 [astro-ph.CO]].


\bibitem{KM}
  B.~Holdom,
  Phys.\ Lett.\  {\bf 166B}, 196 (1986) and
  Phys.\ Lett.\ B {\bf 178}, 65 (1986); 
  K.~R.~Dienes, C.~F.~Kolda and J.~March-Russell,
  Nucl.\ Phys.\ B {\bf 492}, 104 (1997)
  [hep-ph/9610479];
  F.~Del Aguila,
  Acta Phys.\ Polon.\ B {\bf 25}, 1317 (1994)
  [hep-ph/9404323];
  K.~S.~Babu, C.~F.~Kolda and J.~March-Russell,
  Phys.\ Rev.\ D {\bf 54}, 4635 (1996)
  [hep-ph/9603212];
  T.~G.~Rizzo,
  Phys.\ Rev.\ D {\bf 59}, 015020 (1998)
  [hep-ph/9806397].

\bibitem{vectorportal} 
 There has been a huge amount of historical work on this subject; see, for example, 
  D.~Feldman, B.~Kors and P.~Nath,
  Phys.\ Rev.\ D {\bf 75}, 023503 (2007)
  [hep-ph/0610133];
  D.~Feldman, Z.~Liu and P.~Nath,
  Phys.\ Rev.\ D {\bf 75}, 115001 (2007)
  [hep-ph/0702123 [HEP-PH]].;
  M.~Pospelov, A.~Ritz and M.~B.~Voloshin,
  Phys.\ Lett.\ B {\bf 662}, 53 (2008)
  [arXiv:0711.4866 [hep-ph]];
  M.~Pospelov,
  Phys.\ Rev.\ D {\bf 80}, 095002 (2009)
  [arXiv:0811.1030 [hep-ph]]; 
  H.~Davoudiasl, H.~S.~Lee and W.~J.~Marciano,
  Phys.\ Rev.\ Lett.\  {\bf 109}, 031802 (2012)
  [arXiv:1205.2709 [hep-ph]] and 
  Phys.\ Rev.\ D {\bf 85}, 115019 (2012)
  [arXiv:1203.2947 [hep-ph]];
  R.~Essig {\it et al.},
  arXiv:1311.0029 [hep-ph];
  E.~Izaguirre, G.~Krnjaic, P.~Schuster and N.~Toro,
  Phys.\ Rev.\ Lett.\  {\bf 115}, no. 25, 251301 (2015)
  [arXiv:1505.00011 [hep-ph]];
  M.~Khlopov,
  Int.\ J.\ Mod.\ Phys.\ A {\bf 28}, 1330042 (2013)
  [arXiv:1311.2468 [astro-ph.CO]];
 For a general overview and introduction to this framework, see  
  D.~Curtin, R.~Essig, S.~Gori and J.~Shelton,
  JHEP {\bf 1502}, 157 (2015)
  [arXiv:1412.0018 [hep-ph]].

\bibitem{Gherghetta:2019coi}
T.~Gherghetta, J.~Kersten, K.~Olive and M.~Pospelov,
Phys. Rev. D \textbf{100}, no.9, 095001 (2019)
[arXiv:1909.00696 [hep-ph]].

 
\bibitem{Fabbrichesi:2020wbt}
M.~Fabbrichesi, E.~Gabrielli and G.~Lanfranchi,
[arXiv:2005.01515 [hep-ph]].

\bibitem{Graham:2021ggy}
M.~Graham, C.~Hearty and M.~Williams,
[arXiv:2104.10280 [hep-ph]].

\bibitem{Barducci:2021egn}
D.~Barducci, E.~Bertuzzo, G.~Grilli di Cortona and G.~M.~Salla,
JHEP \textbf{12}, 081 (2021)
[arXiv:2109.04852 [hep-ph]].

\bibitem{Alonso-Gonzalez:2025xqg}
D.~Alonso-Gonz{\'a}lez, D.~Cerde{\~n}o, P.~Foldenauer and J.~M.~No,
[arXiv:2507.11376 [hep-ph]].

\bibitem{Jorge:2026bbs}
A.~W.~R.~Jorge, L.~Sagunski, G.~W.~Yuan, T.~Song and E.~Bratkovskaya,
[arXiv:2601.15066 [hep-ph]].

\bibitem{Caputo:2026pdw}
A.~Caputo and R.~Essig,
[arXiv:2603.08430 [hep-ph]].

\bibitem{Li:2024wqj}
S.~Li, J.~M.~Yang, M.~Zhang and R.~Zhu,
[arXiv:2405.18226 [hep-ph]].


\bibitem{Rizzo:2018vlb}
T.~G.~Rizzo,
Phys. Rev. D \textbf{99}, no.11, 115024 (2019)
[arXiv:1810.07531 [hep-ph]].

\bibitem{Rueter:2019wdf}
T.~D.~Rueter and T.~G.~Rizzo,
Phys. Rev. D \textbf{101}, no.1, 015014 (2020)
[arXiv:1909.09160 [hep-ph]].

\bibitem{Kim:2019oyh}
J.~H.~Kim, S.~D.~Lane, H.~S.~Lee, I.~M.~Lewis and M.~Sullivan,
Phys. Rev. D \textbf{101}, no.3, 035041 (2020)
[arXiv:1904.05893 [hep-ph]].

\bibitem{Rueter:2020qhf}
T.~D.~Rueter and T.~G.~Rizzo,
[arXiv:2011.03529 [hep-ph]].

\bibitem{Wojcik:2020wgm}
G.~N.~Wojcik and T.~G.~Rizzo,
Phys. Rev. D \textbf{105}, no.1, 015032 (2022)
[arXiv:2012.05406 [hep-ph]].

\bibitem{Rizzo:2021lob}
T.~G.~Rizzo,
JHEP \textbf{11}, 035 (2021)
[arXiv:2106.11150 [hep-ph]].

\bibitem{Rizzo:2022qan}
T.~G.~Rizzo,
[arXiv:2202.02222 [hep-ph]].

\bibitem{Wojcik:2022rtk}
G.~N.~Wojcik,
[arXiv:2205.11545 [hep-ph]].

\bibitem{Rizzo:2022jti}
T.~G.~Rizzo,
Phys. Rev. D \textbf{106}, no.3, 035024 (2022)
[arXiv:2206.09814 [hep-ph]].

\bibitem{Rizzo:2022lpm}
T.~G.~Rizzo,
Phys. Rev. D \textbf{106}, no.9, 095024 (2022)
[arXiv:2209.00688 [hep-ph]].

\bibitem{Carvunis:2022yur}
A.~Carvunis, N.~McGinnis and D.~E.~Morrissey,
[arXiv:2209.14305 [hep-ph]].

\bibitem{Verma:2022nyd}
S.~Verma, S.~Biswas, A.~Chatterjee and J.~Ganguly,
[arXiv:2209.13888 [hep-ph]].

\bibitem{Rizzo:2023qbj}
T.~G.~Rizzo,
Phys. Rev. D \textbf{107}, no.9, 095014 (2023)
[arXiv:2302.12698 [hep-ph]].

\bibitem{Bauer:2022nwt}
M.~Bauer and P.~Foldenauer,
Phys. Rev. Lett. \textbf{129}, no.17, 171801 (2022)
[arXiv:2207.00023 [hep-ph]].


\bibitem{Wojcik:2022woa}
G.~N.~Wojcik, L.~L.~Everett, S.~T.~Eu and R.~Ximenes,
Phys. Lett. B \textbf{841}, 137931 (2023)
[arXiv:2211.09918 [hep-ph]].

\bibitem{Wojcik:2023ggt}
G.~N.~Wojcik, L.~L.~Everett, S.~T.~Eu and R.~Ximenes,
Phys. Rev. D \textbf{108}, no.5, 055033 (2023)
[arXiv:2303.12983 [hep-ph]].


\bibitem{Rizzo:2023kvy}
T.~G.~Rizzo,
Phys. Rev. D \textbf{108}, no.5, 055021 (2023)
[arXiv:2307.08508 [hep-ph]].

\bibitem{Rizzo:2023djp}
T.~G.~Rizzo,
Phys. Rev. D \textbf{109}, no.5, 055039 (2024)
[arXiv:2312.00226 [hep-ph]].

\bibitem{Rizzo:2024bhn}
T.~G.~Rizzo,
Phys. Rev. D \textbf{110}, no.7, 075037 (2024)
[arXiv:2408.01296 [hep-ph]].

\bibitem{Ardu:2024bxg}
M.~Ardu, M.~H.~Rahat, N.~Valori and O.~Vives,
[arXiv:2407.21100 [hep-ph]].

\bibitem{Rizzo:2024kzu}
T.~G.~Rizzo,
Phys. Rev. D \textbf{111}, no.7, 075018 (2025)
[arXiv:2412.17174 [hep-ph]].

\bibitem{Rizzo:2025tap}
T.~G.~Rizzo,
Phys. Rev. D \textbf{112}, no.1, 015029 (2025)
doi:10.1103/h81b-ydmw
[arXiv:2505.04474 [hep-ph]].

\bibitem{Tewary:2025vij}
K.~Tewary, S.~Biswas and S.~Verma,
[arXiv:2511.00578 [hep-ph]].

\bibitem{Rizzo:2025fsy}
T.~G.~Rizzo,
[arXiv:2511.13632 [hep-ph]].

\bibitem{Rizzo:2026dsj}
T.~G.~Rizzo,
[arXiv:2602.24122 [hep-ph]].

\bibitem{Rizzo:2026mza}
T.~G.~Rizzo,
[arXiv:2607.15965 [hep-ph]].

\bibitem{CarcamoHernandez:2023wzf}
For a recent review of vector-like fermions, see 
A.~E.~C\'arcamo Hern\'andez, K.~Kowalska, H.~Lee and D.~Rizzo,
[arXiv:2309.13968 [hep-ph]].

\bibitem{CMS:2024bni}
A.~Hayrapetyan \textit{et al.} [CMS],
[arXiv:2405.17605 [hep-ex]].

\bibitem{Alves:2023ufm}
J.~M.~Alves, G.~C.~Branco, A.~L.~Cherchiglia, C.~C.~Nishi, J.~T.~Penedo, P.~M.~F.~Pereira, M.~N.~Rebelo and J.~I.~Silva-Marcos,
Phys. Rept. \textbf{1057}, 1-69 (2024)
[arXiv:2304.10561 [hep-ph]].

\bibitem{Banerjee:2024zvg}
A.~Banerjee, E.~Bergeaas Kuutmann, V.~Ellajosyula, R.~Enberg, G.~Ferretti and L.~Panizzi,
[arXiv:2406.09193 [hep-ph]].

\bibitem{Guedes:2021oqx}
G.~Guedes and J.~Santiago,
JHEP \textbf{01}, 111 (2022)
[arXiv:2107.03429 [hep-ph]].

\bibitem{Adhikary:2024esf}
A.~Adhikary, M.~Olechowski, J.~Rosiek and M.~Ryczkowski,
[arXiv:2406.16050 [hep-ph]].

\bibitem{Benbrik:2024fku}
R.~Benbrik, M.~Boukidi, M.~Ech-chaouy, S.~Moretti, K.~Salime and Q.~S.~Yan,
[arXiv:2412.01761 [hep-ph]].

\bibitem{Albergaria:2024pji}
F.~Albergaria, J.~F.~Bastos, B.~Belfatto, G.~C.~Branco, J.~T.~Penedo, A.~Rodr\'\i{}guez-S\'anchez and J.~I.~Silva-Marcos,
[arXiv:2412.21201 [hep-ph]].

\bibitem{Chen:2017hak}
C.~Y.~Chen, S.~Dawson and E.~Furlan,
Phys. Rev. D \textbf{96}, no.1, 015006 (2017)
[arXiv:1703.06134 [hep-ph]].

\bibitem{Biekotter:2016kgi}
A.~Biek{\"o}tter, J.~L.~Hewett, J.~S.~Kim, M.~Kr{\"a}mer, T.~G.~Rizzo, K.~Rolbiecki, J.~Tattersall and T.~Weber,
Int. J. Mod. Phys. A \textbf{32}, no.05, 1750032 (2017)
[arXiv:1608.01312 [hep-ph]].


\bibitem{Arcadi:2017kky} 
 G.~Arcadi, M.~Dutra, P.~Ghosh, M.~Lindner, Y.~Mambrini, M.~Pierre, S.~Profumo and F.~S.~Queiroz, 
Eur. Phys. J. C \textbf{78}, no.3, 203 (2018)
[arXiv:1703.07364 [hep-ph]].
  
\bibitem{Roszkowski:2017nbc}
L.~Roszkowski, E.~M.~Sessolo and S.~Trojanowski,
Rept. Prog. Phys. \textbf{81}, no.6, 066201 (2018)
[arXiv:1707.06277 [hep-ph]].

\bibitem{Arcadi:2024ukq}
G.~Arcadi, D.~Cabo-Almeida, M.~Dutra, P.~Ghosh, M.~Lindner, Y.~Mambrini, J.~P.~Neto, M.~Pierre, S.~Profumo and F.~S.~Queiroz,
[arXiv:2403.15860 [hep-ph]].
 
\bibitem{Hewett:1988xc}
J.~L.~Hewett and T.~G.~Rizzo,
Phys. Rept. \textbf{183}, 193 (1989). 



\bibitem{limit}
The most up to date determinations of the current bound on $\Gamma_{inv}$ and the expectations for Tera-Z sensitivity to this parameter have been presented in: 
A. ~Freitas, talk given at LoopFest XXIV, Brookhaven National Laboratory, Upton, NY, 27-29 May, 2026 and P.~Meade, talk given at the Biennial Meeting of the Division of Particles and
Fields, Fermilab, Batavia, IL, 20-24 July, 2026. See also,
F. Takahashi et al. (Particle Data Group), Int. J. Mod. Phys. A 41 , 2630011 (2026); 
Y.~Fischer \textit{et al.} [Gfitter Group],
[arXiv:2607.09861 [hep-ph]]; 
P.~Olgoso, P.~Paradisi and N.~Selimovic,
JHEP \textbf{04}, 106 (2026)
[arXiv:2507.17803 [hep-ph]]; 
V.~Maura, B.~A.~Stefanek and T.~You,
JHEP \textbf{10}, 022 (2025)
[arXiv:2412.14241 [hep-ph]];
M.~Selvaggi \textit{et al.} [FCC],
J.~Erler and M.~Schott,
CERN Yellow Reports: Monographs \textbf{3}, 217-219 (2020);
M.~Drewes, J.~Klari{\'c} and Y.~Z.~Li,
Eur. Phys. J. C \textbf{86}, no.4, 337 (2026)
[arXiv:2511.23461 [hep-ph]].



\bibitem{bj}
J. Bjorken, in Proceedings of the 1976 SLAC Summer Institute on Particle Physics, ed. by M.C. Zipf (SLAC Report 198, 1977) p. 1.

\bibitem{Invisible}
 ATLAS Collaboration, 
``Combination of searches for invisible Higgs boson decays with the ATLAS experiment,''
ATLAS-CONF-2020-052;
G.~Aad \textit{et al.} [ATLAS],
Phys. Lett. B \textbf{842}, 137963 (2023)
[arXiv:2301.10731 [hep-ex]];
V.~Milosevic [CMS],
PoS \textbf{EPS-HEP2021}, 602 (2022)


\bibitem{hoomie}
H.~Davoudiasl, H.~S.~Lee and W.~J.~Marciano,
Phys. Rev. D \textbf{85}, 115019 (2012)
[arXiv:1203.2947 [hep-ph]];
H.~Davoudiasl, H.~S.~Lee and W.~J.~Marciano,
Phys. Rev. D \textbf{86}, 095009 (2012)
[arXiv:1208.2973 [hep-ph]].



\bibitem{GBET}
  M.~S.~Chanowitz and M.~K.~Gaillard,
  Nucl.\ Phys.\ B {\bf 261}, 379 (1985);
  B.~W.~Lee, C.~Quigg and H.~B.~Thacker,
  Phys.\ Rev.\ D {\bf 16}, 1519 (1977);
  J.~M.~Cornwall, D.~N.~Levin and G.~Tiktopoulos,
  Phys.\ Rev.\ D {\bf 10}, 1145 (1974)
  Erratum: [Phys.\ Rev.\ D {\bf 11}, 972 (1975)];
  G.~J.~Gounaris, R.~Kogerler and H.~Neufeld,
  Phys.\ Rev.\ D {\bf 34}, 3257 (1986).


\bibitem{next}
T.G.~Rizzo, work in progress. 





\end{thebibliography}
\end{document}